\documentclass[10pt,aps,prb,twocolumn,superscriptaddress, nolongbibliography]{revtex4-2}

\usepackage{lmodern}
\usepackage[T1]{fontenc}
\setcitestyle{super}

\usepackage[utf8]{inputenc}
\usepackage{graphicx}
\usepackage{xcolor}

\usepackage[colorlinks=true,bookmarks=false,citecolor=blue,urlcolor=blue,linkcolor=red]{hyperref}

\begin{document}

\title{Femtojoule, femtosecond all-optical switching in lithium niobate nanophotonics}

\author{
Qiushi Guo$^{1,\ast}$, Ryoto Sekine$^{1,\ast}$, Luis Ledezma$^{1,2,\ast}$, Rajveer Nehra$^1$, Devin J. Dean$^3$, Arkadev Roy$^1$, Robert M. Gray$^1$, Saman Jahani$^1$ and Alireza Marandi$^{1,\dagger}$\\
\textit{
$^1$Department of Electrical Engineering, California Institute of Technology, Pasadena, California 91125, USA. \\
$^2$Jet Propulsion Laboratory, California Institute of Technology, Pasadena, California 91109, USA.\\
$^3$Department of Applied Physics, Cornell University, Ithaca, New York 14850, USA.\\
$^\ast$These authors contributed equally to this work.}\\
$^\dagger$Email: \href{mailto:marandi@caltech.edu}{marandi@caltech.edu}
}

\date{\today}

\maketitle

\textbf{
Optical nonlinear functions are crucial for various applications in integrated photonics, such as all-optical information processing, photonic neural networks and on-chip ultrafast light sources. Due to the weak nonlinearities in most integrated photonic platforms, realizing optical nonlinear functions typically requires large driving energies in the picojoules level or beyond, thus imposing a barrier for most applications. Here, we tackle this challenge and demonstrate an integrated nonlinear splitter device in lithium niobate nano-waveguides by simultaneous engineering of the dispersion and quasi-phase matching. We achieve non-resonant all-optical switching with ultra-low energies down to tens of femtojoules, a near instantaneous switching time of 18 fs, and a large extinction ratio of more than 5 dB. Our nonlinear splitter simultaneously realizes switch-on and -off operations and features a state-of-the-art switching energy-time product as low as \boldmath{$1.4 \times10^{-27}$} J$\cdot$s. We also show a path toward attojoule level all-optical switching by further optimizing the device geometry. Our results can enable on-chip ultrafast and energy-efficient all-optical information processing, computing systems, and light sources. 
}

\begin{figure*}[ht]
\centering
\includegraphics[width=0.9\linewidth]{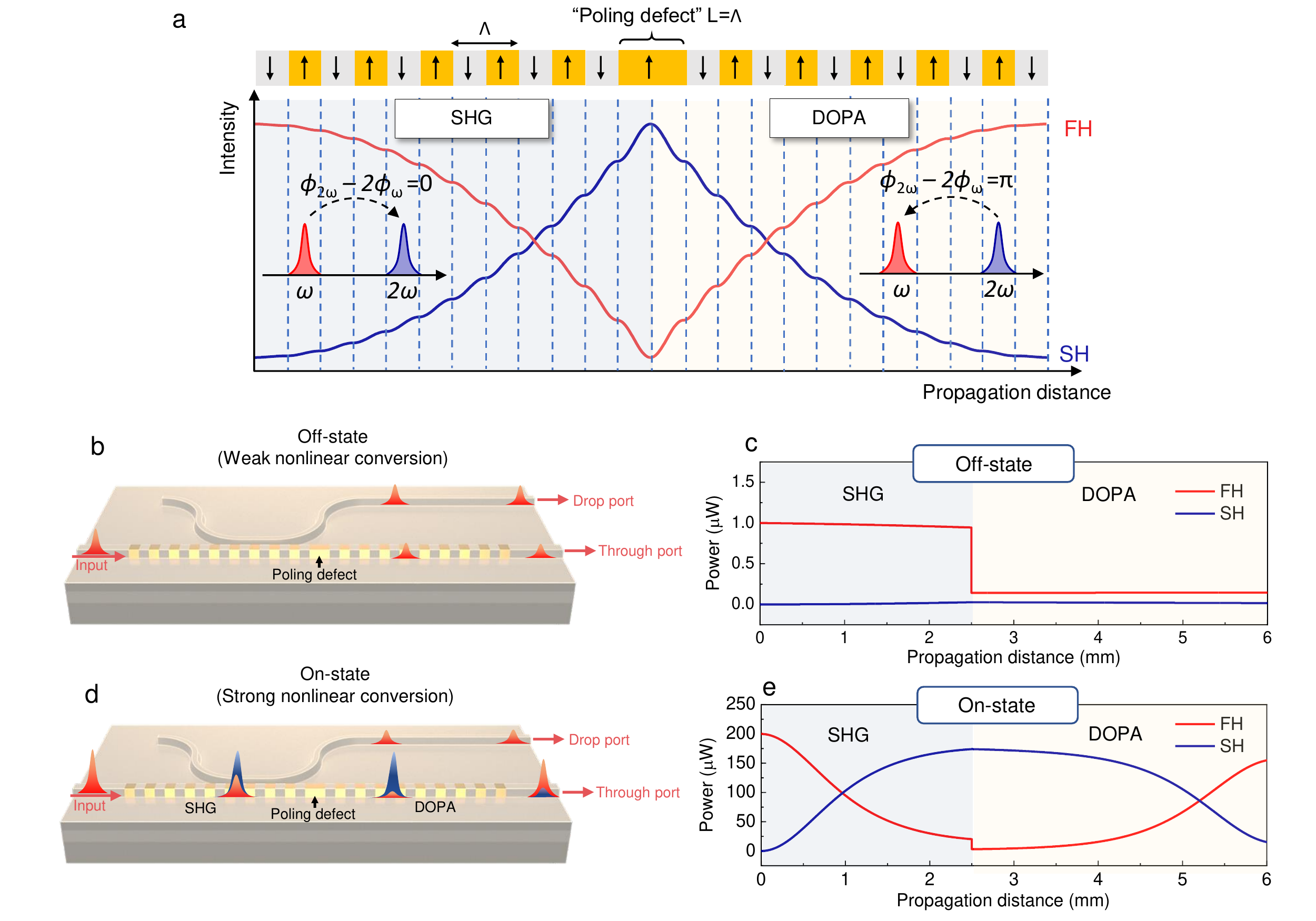}
\caption{\textbf{Device design and operating principle.} 
\textbf{a}, Illustration of the QPM engineering in PPLN waveguide. The poling defect (a longer ferroelectric domain with a length $L=\Lambda$) shifts the phase difference between the FH and SH by $\pi$. As a result, the poling defect switches the SHG to degenerate optical parametric amplification (DOPA) in the second half of the PPLN waveguide. The insets illustrate the phase relation and the direction of energy transfer between the FH and the SH. \textbf{b}, Schematic of the integrated nonlinear splitter device and its operation in the ``off-state'' when the input FH pulse energy is low. \textbf{c}, Simulated evolution of the FH and SH optical power along the main waveguide for the off-state, in which the transmittance of the FH is low (15$\%$). \textbf{d}, Schematic of the integrated nonlinear splitter device and its operation in the ``on-state'' when the input pulse energy is high. \textbf{e}, Simulated evolution of the FH and SH optical power along the main waveguide for the ``on-state'', in which the transmittance of the FH is high (85$\%$). The simulations in \textbf{c} and \textbf{e} assume an input pulse of 46.2 fs at 2.09 $\mu$m and device parameters of a 2.5-mm-long SHG region, a 3.5-mm-long DOPA region, and 85$\%$ out-coupling of the FH by the directional coupler.
}\label{Fig1}
\end{figure*}

Photons are known to be excellent information carriers. Yet, the quest for all-optical information processing -- a technology that can potentially eliminate the limitations on the bandwidth and energy consumption of electronic and opto-electronic systems -- is generally deemed to be challenging because optical nonlinearities are usually weak. All-optical switching using cubic ($\chi^{(3)}$)  nonlinearities\cite{grinblat2019ultrafast,yang2017femtosecond} and saturable absorption\cite{iizuka2006all,takahashi1996ultrafast,spuhler2005semiconductor} in semiconductor materials typically require pulse energies on the order of picojoules or beyond. Such  energy requirements hinder their widespread utilization for any applications as they necessiate bulky and power-hungry light sources. To lower the energy requirement of all-optical switching, one approach consists of enhancing the optical nonlinearity in optical cavities. However, this enhancement is accompanied by a cavity photon lifetime, which unavoidably increases the switching times and typically leads to low bandwidths up to 10s of GHz\cite{nozaki2010sub,almeida2004all,tanabe2007fast,hu2008picosecond,martinez2010ultrafast,yanik2003all,chai2017ultrafast}. Therefore, all-optical switching in solid-state photonic platforms faces a performance trade-off between the energy per bit and the switching time, making the energy-time product an apt figure of merit\cite{ono2020ultrafast}. A promising path toward a better energy-time product is the utilization of stronger and instantaneous nonlinearities, such as the quadratic nonlinearity ($\chi^{(2)}$).

Compared to the $\chi^{(3)}$ nonlinearity, $\chi^{(2)}$ nonlinearity in non-centrosymmetric materials requires lower light intensities by many orders of magnitude. When the phase-matching condition is achieved and in the absence of significant dispersion, nonlinear optical interactions can grow substantially as a function of propagation length, thus circumventing the need of resonant enhancement and compromising switching speed. Recently, thin-film lithium niobate (TFLN) has emerged as a promising integrated photonic platform. TFLN-based nanophotonic waveguides exhibit strong $\chi^{(2)}$ nonlinearity and a high normalized nonlinear frequency conversion efficiency in the continuous wave (CW) regime ($>$1000$\%$/W-cm$^2$)\cite{wang2018ultrahigh,zhao2020shallow,chen2019ultra,rao2019actively} that is not easily attainable in bulk material platforms, thanks to the strong spatial confinement of the waveguide modes as well as quasi-phase matching (QPM). Additionally, the tight spatial confinement of the waveguide modes allows dispersion engineering\cite{jankowski_ultrabroadband_2020,ledezma2021intense,jankowski2021efficient}, which enables temporal confinement of interacting waves over long interaction lengths, leading to further enhancement of the nonlinear processes using ultra-short pulses. 

In this work, we engineer the quasi-phase matching in a TFLN nanophotonic waveguide to achieve non-resonant all-optical switching in an integrated nonlinear splitter. Simultaneously, dispersion engineering of our device allows spatio-temporal confinement of the pulses over the length of the device, leading to ultra-low energy (fJ) and ultrafast (fs) all-optical switching with a high exinction ratio. The dispersion- and QPM- engineered LN nanophotonic waveguide enables nonlinear interaction lengths as long as 70 mm, allowing for attojoule ultrafast all-optical switching.

The main element of our all-optical switch is a QPM-engineered LN nanophotonic waveguide. Figure 1 illustrates the concept of this QPM-engineering: a uniform periodically poled LN waveguide (periodicity=$\Lambda$) is perturbed by a ‘‘poling defect’’, i.e. an isolated ferroelectric domain of length $L=\Lambda$ in the middle of the waveguide. While the poling period is designed for phase-matched second-harmonic generation (SHG), the poling defect locally changes the phase relationship between the first harmonic (FH) and the second harmonic (SH) waves by the amount of $\Delta\phi=\pi$\cite{gallo2001all}. Since the direction of power flow between the FH and the SH is dependent on the relative phase between them, the $\pi$ phase shift due to the poling defect switches the nonlinear process from SHG to degenerate optical parametric amplification (DOPA), in which the generated SH serves as the pump to amplify the FH.
\begin{figure*}[ht]
\centering
\includegraphics[width=1.01\linewidth]{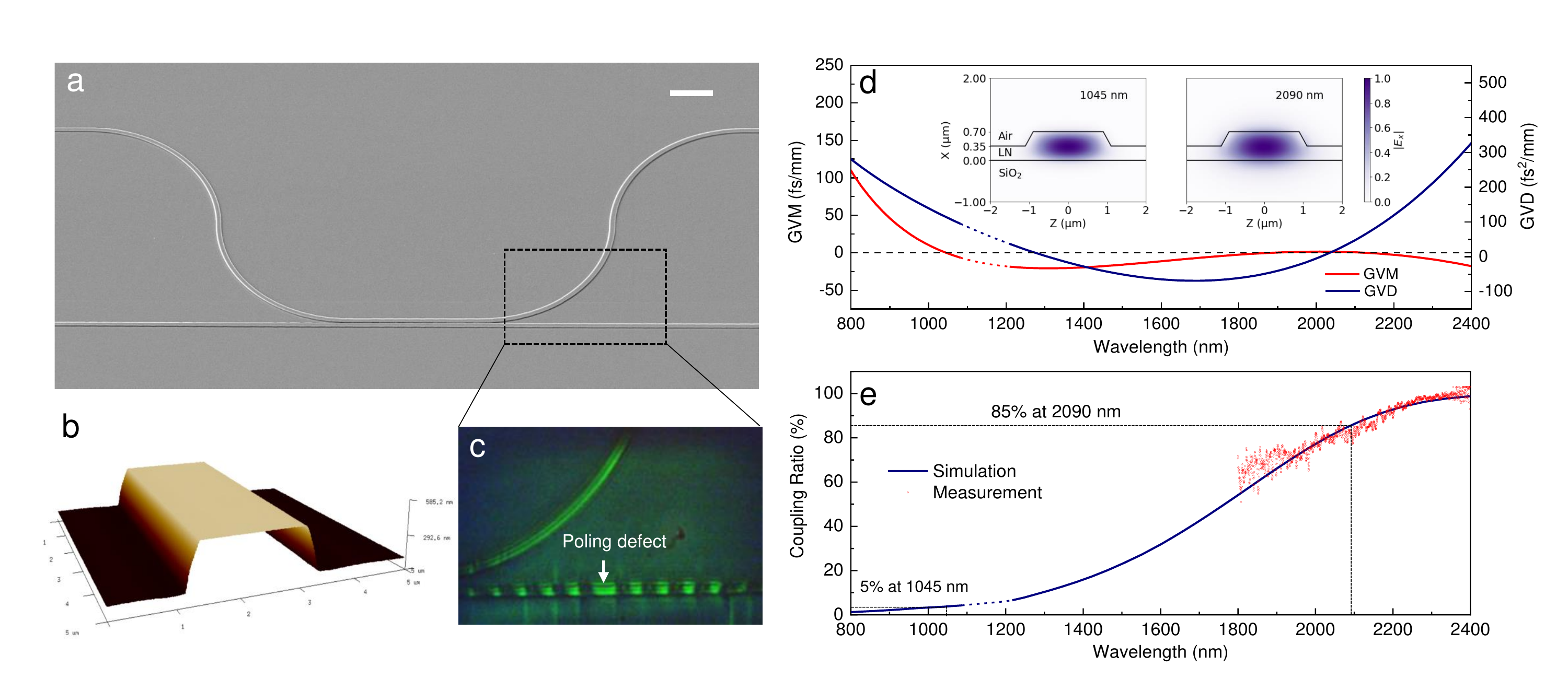}
\caption{\textbf{Integrated nonlinear splitter and its linear optical characteristics.}
\textbf{a}, SEM image of the fabricated nonlinear splitter. Scale bar: 20 $\mu$m. The device has a 2.5-mm-long SHG region and a 3.5-mm-long DOPA region. The directional coupler has a coupling length of 70 $\mu$m and a gap of 650 nm between the waveguide top surfaces. \textbf{b}, 2D AFM scan on LN waveguide.  \textbf{c}, Second harmonic microscope image showing the inverted domains and the poling defect along the waveguides. \textbf{d}, Simulated group velocity mismatch (GVM, red) between the FH and the SH and group velocity dispersion (GVD, blue) for the quasi-TE modes of the dispersion-engineered LN waveguide. The optimized waveguide has a top width of 1650 nm, an etching depth of 350 nm and a total thin-film thickness of 700 nm. The waveguide exhibits low (0.4 fs/mm) GVM between the pump at 1045 nm and the signal around 2090 nm, and low GVD for both wavelengths. Inset: Electric field distributions of the fundamental quasi-TE modes for the dispersion-engineered waveguide at 1045 nm and 2090 nm. The black dashed line denotes the zero GVM. \textbf{e}, Measured (red symbols) and simulated (blue solid curve) out-coupling ratio of the directional coupler as a function of the wavelength. In \textbf{d} and \textbf{e}, the dotted lines correspond to the regime where mode crossing between the fundamental TE mode and the second order TM mode occurs, which is distant enough from the SH central wavelength and is not expected to affect the device operation.
}\label{Fig2}
\end{figure*}

In addition to the QPM-engineered main waveguide, our all-optical switch is composed of a neighboring linear directional coupler, as sketched in Fig. 1b and d. The linear directional coupler evenescently couples out most (85$\%$) of the FH, while leaving most of SH freely propagating in the main waveguide. This whole device exhibits a strongly intensity-dependent
splitting ratio. When the input FH intensity is low (in the “off-state'' shown in Fig. 1b), most of the input FH does not convert to SH, and hence is directed by the linear coupler to the drop port. This is illustrated by the simulated power evolution of both the FH and the SH in the main waveguide in Fig. 1c. In this “off-state'', the transmittance of the FH in the main waveguide (through port) is low. However, when the input FH intensity is high (or in the “on-state'' as shown in Fig. 1d, e), due to the efficient SHG at the beginning of the waveguide, most of the FH can convert to the SH, which remains in the main waveguide after passing through the coupler, and negligible FH is directed to the drop port. In the second half of the main waveguide, the poling defect switches the SHG process to the DOPA, through which most of the SH converts back to the FH. As shown in Fig. 1e, in the ``on-state'', the device favors transmission of the FH to the through port since most of input pulse energy can be ``stored'' in (i.e. converted to) the SH, which is unaffected by the linear coupler. Since the FH transmission strongly depends on the input pulse energy of the FH, the intensity-dependent nonlinear splitter functions as an all-optical 
switch.

\begin{figure*}[ht]
\centering
\includegraphics[width=0.95\linewidth]{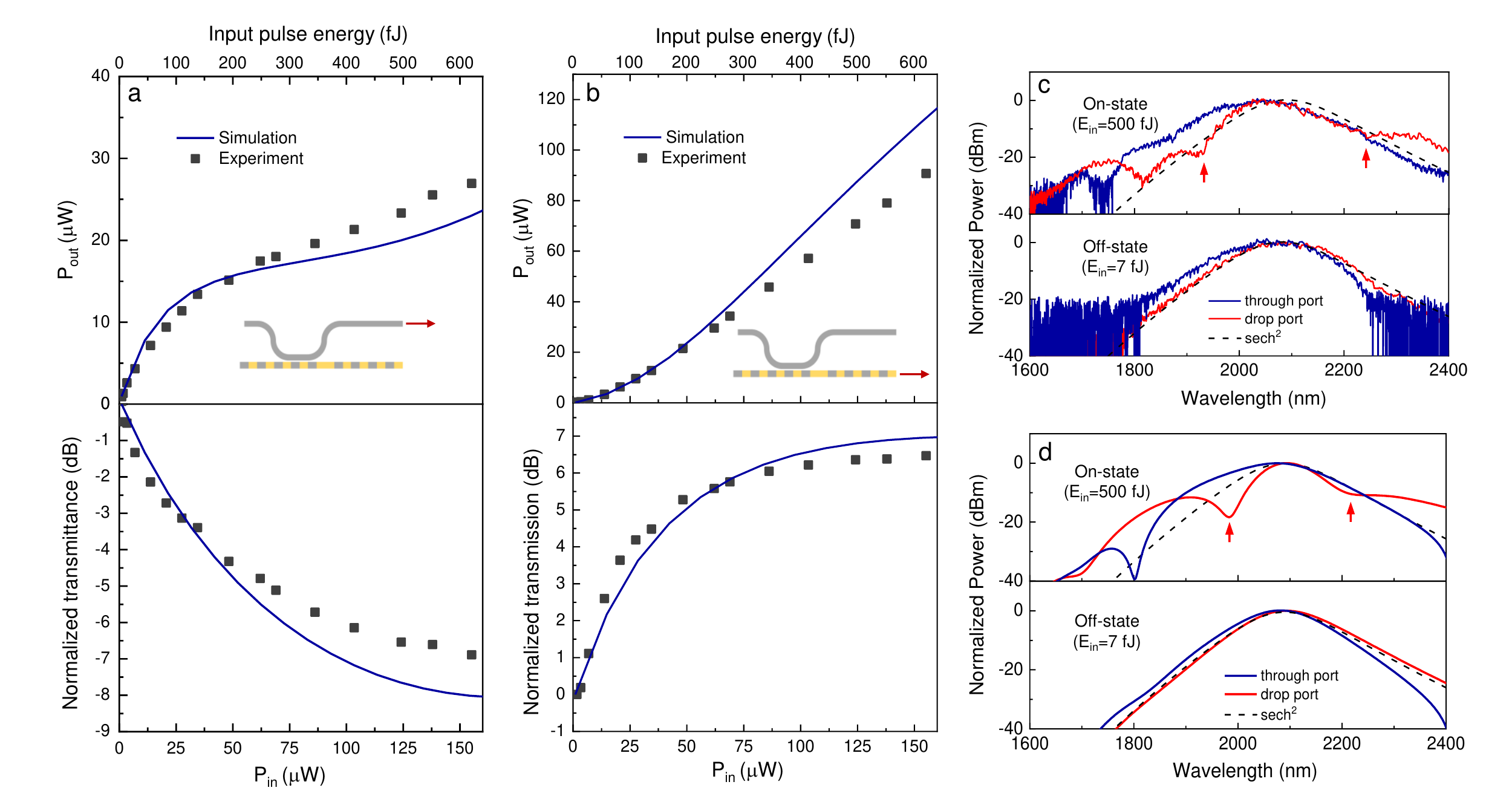}
\caption{\textbf{Ultra-low energy nonlinear optical transmission in the integrated nonlinear splitter.} 
\textbf{a}, Upper panel: average on-chip output power of 2.09 $\mu$m FH from the drop port as a function of on-chip input average power/pulse energy. Lower panel: normalized transmittance ($10\log(T/T_{P_\mathrm{in}\rightarrow0})$) of the FH from the drop port. \textbf{b}, Upper panel: average output power of FH from the through port as a function of input average power/pulse energy. Lower panel: normalized transmittance ($10\log(T/T_{P_\mathrm{in}\rightarrow0})$) of the FH from the through port. In both \textbf{a} and \textbf{b}, the blue solid lines are the simulation results. The black symbols are the measured data. The insets illustrate the ports at which we collected the data. \textbf{c}, Measured output FH spectra at the through port (blue) and the drop port (red) in the ``on-state'' (upper panel) and the ``off-state'' (lower panel). The input FH pulse energy is 500 fJ for the ``on-state'' and 7 fJ for the ``off-state''. In the ``on-state'', the dip in the spectrum of the drop port, which does not persist in the through port or in the ``off-state'', is another strong experimental evidence for the interplay of SHG and DOPA in the switching mechanism. \textbf{d}, Simulated output FH spectra at the through port (blue) and the drop port (red) in the ``on-state'' (upper panel) and the ``off-state'' (lower panel), corresponding to the measurements in (c). In both (c) and (d), the spectral dips are labeled by the red arrows.
}\label{Fig3}
\end{figure*}

It is worth noting that our design is in sharp contrast with the previously demonstrated nonlinear directional couplers in micro-wavguides which required picojoules of energy\cite{schiek2012temporal}. In those devices, quadratic nonlinearity perturbs the evanescent coupling as the switching mechanism, while in our device the switching mechanism arises from the wavelength conversion in the QPM-engineered waveguide and the evanescent coupler remains a linear spectrally-selective splitter. Our nonlinear splitter shares similarities in operation with bulk nonlinear mirrors used in some mode-locked lasers\cite{stankov1988mirror}.

We fabricated the nonlinear splitter on a 700-nm-thick X-cut magnesium-oxide (MgO) doped LN thin film on a 2-µm-thick silicon dioxide layer on top of a LN substrate (NANOLN). The details about the periodic poling, the waveguide patterning and etching can be found in the Methods. As shown in the scanning electron microscope image (Fig. 2a) and the atomic force microscope image (Fig. 2b), the Ar-based dry etching process yields smooth waveguide sidewalls (with surface roughness $\sim 1$ nm) and a sidewall angle of $\sim 60^{\circ}$. The top width of the main waveguide and the etching depth are measured to be 1650 nm and 350 nm, respectively, with an error of $\pm 5$ nm. The inverted ferroelectric domains and the poling defect along the main waveguide are shown in the second-harmonic image in Fig. 2c. The device has a 2.5-mm-long SHG region and a 3.5-mm-long DOPA region.

To achieve ultra-low-energy and ultrafast operation, we engineer the dispersion of the LN ridge waveguide and minimize both the group velocity dispersion (GVD) and the group velocity mismatch (GVM) between the FH and the SH\cite{ledezma2021intense}. Negligible GVD at the FH and SH wavelengths preserves the temporal confinement of these pulses and hence their high peak intensities along the waveguide, thereby ensuring efficient short-pulse and low-energy SHG and DOPA. Additionally, negligible GVM between the FH and the SH waves guarantees that both FH and SH pulses travel together along the waveguide. As shown in Fig. 2d, in the dispersion-engineered waveguide, the fundamental quasi-TE modes at the FH (2090 nm) and SH (1045 nm) have a very low GVM of 0.4 fs/mm. In addition, the GVD for both the FH and SH waves are as low as 40 fs$^2$/mm and 114 fs$^2$/mm, respectively. For a 35-fs-long input pulse at 2.09 $\mu$m, the optimized waveguide has a dispersion length of more than 50 mm and a walk-off length of 115 mm. 

To ensure that the coupling ratio of the directional coupler is resilient to fabrication errors, we adopt an adiabatic design in which the main waveguide is uniform with a fixed width, while the coupler waveguide width is adiabatically tapered\cite{guo201670}. The detailed geometry of the directional coupler is discussed in the Supplementary Information Section I. Figure 2e shows the wavelength dependent coupling ratio, which is the ratio between the output power at the drop port over the input power. Due to the large mode area difference between the fundemental TE modes at 2090 nm and 1045 nm, the 70-$\mu$m-long directional coupler exhibits a large coupling ratio of over 85$\%$ for wavelengths beyond 2090 nm and a small coupling ratio of less than 5$\%$ for wavelengths below 1045 nm. The measured (red symbols) result agrees well with the simulation result (blue solid curve). 

We characterized the nonlinear optical behavior of the device using 46-fs-long pulses at 2.09 µm from a synchronously pumped degenerate optical parametric oscillator (OPO) with a repetition frequency of 250 MHz. The characterization of the 2.09 µm pulses is elaborated in Supplementary Information Section II.  By measuring the nonlinear splitter devices with different poling periods, we found that 5.11 $\mu$m is the optimal poling period for realizing QPM, which is in good agreement with the theoretical period of 5.08 $\mu$m. The experimental details of optimizing the QPM condition are discussed in Methods and the Supplementary Information Section IV. At the optimum QPM condition, we measured the output power of the FH both at the through port and the drop port. The detailed calibration of the input/outout coupling loss of our waveguide is discussed in the Supplementary Information Section V.

As shown in Fig. 3a, the normalized transmittance from the input to the drop port, defined as $10\log(T/T_{P_\mathrm{in}\rightarrow0})$, shows a clear reduction (\textasciitilde 7 dB) when the on-chip input FH pulse energy increases from 0 to 600 fJ. This behavior is well captured by the simulation (blue solid line). Such a reduction in the transmittance is a result of the strong depletion of FH waves during the SHG process, since the directional coupler only couples out the FH in the first half of the main waveguide. The time-domain nonlinear dynamics of the SHG process is elaborated in the Supplementary Information Section VI. Based on the measured depletion of the FH shown in Fig. 3a and assuming that the propagation loss of the FH is far less than its depletion, we estimated that 600 fJ input FH pulse energy in the waveguide converts to a SH pulse energy of 500 fJ. 

\begin{figure*}[ht]
\centering
\includegraphics[width=0.85\linewidth]{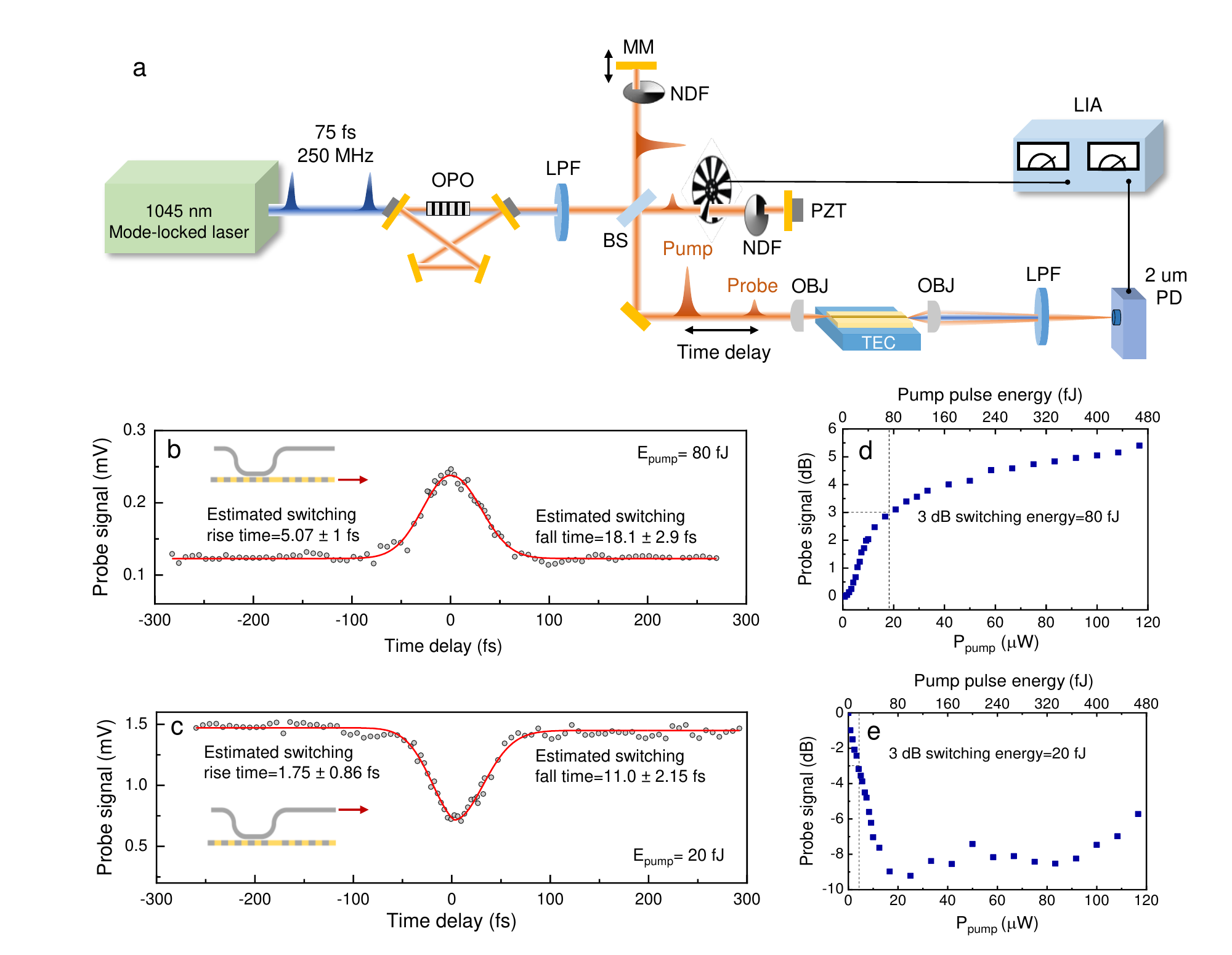}
\caption{\textbf{Femtosecond, femtojoule all-optical switching.} \textbf{a}, Experimental setup for femtosecond all optical switching measurement. ~46-fs pulses at 2.09 $\mu$m generated from a free-space optical parametric oscillator (OPO) was used to characterize the device. A pump and probe field with relative time delay $\Delta\tau$ are injected into the waveguide via the objective lens. Probe transmission depends on whether the two pulses excite the device simultaneously or at different times. LPF: long-pass filter; MM: motorized mirror; PZT: piezoelectric transducer; NDF: neutral density filter; BS: beam splitter; OBJ: objective lens; LIA: lock-in amplifier; PD: photodetector. \textbf{b}, Measured dynamics of the ``on-state'' at the through port. The input pump pulse energy ($E_\mathrm{pump}$) is 80 fJ. \textbf{c}, Measured dynamics of the ``on-state'' at the drop port. The $E_\mathrm{pump}$ is 20 fJ. The insets illustrate the ports at which we collected the data. \textbf{d}, relative probe signal power at the through port as a function of the pump pulse power (energy). \textbf{e}, relative probe signal power at the drop port as a function of the pump pulse power (energy). The 3 dB switching energies at the through port and the drop port are 80 fJ and 20 fJ, respectively.
}\label{Fig4}
\end{figure*}

The normalized transmittance of the FH from the input port to the through port increases by more than 5 dB as the input pulse energy increases, as shown in Fig. 3b. This behavior is the opposite of the transmittance to the drop port, which confirms that the poling defect indeed switches the SHG to the DOPA process in the second half of the main waveguide, thereby converting the generated SH back into the FH. The experimental result agrees well with the simulation (blue solid line), despite showing a slightly lower peak transmittance, which can be ascribed to the slightly lower SHG efficiency in the first half of the device or the imperfect phase shift imposed by the poling defect. Based on the results shown in Fig. 3a and b, we can deduce that the largest parametric gain in the DOPA region is more than 12 dB at 600 fJ input FH pulse energy, which translates to a gain per unit length of 34 dB/cm given that the DOPA region is 3.5-mm-long. Notably, this result is consistent with our recent parametric gain measurement on a similar QPM LN waveguide, which exhibits a gain per unit length of 35 dB/cm pumped with 500 fJ 1.045 $\mu$m pulses\cite{ledezma2021intense}. Moreover, the measured output FH power from the main waveguide agrees well with the simulation. 

The interplay of SHG and DOPA in the all-optical switching mechanism is also evident from the output spectra. In Fig. 3c, we compare the measured output FH spectra at the drop port and the through port. When the device is subjected to 500 fJ of input pulse energy (in the ``on-state''), the output spectrum at the drop port (red) deviates from the initial sech$^2$ spectral shape, showing spectral dips around 1920 nm and 2250 nm which indicate a strong depletion of the FH. However, the output spectrum at the through port (blue) does not show such dips at high input pulse energies and has a spectral shape similar to the sech$^2$, which indicates the recovery of the FH power. When the device is in the ``off-state'' (Fig. 3c lower panel), the output spectra measured at the through port and the drop port both have sech$^2$ shapes, because of the weak nonlinear conversion in the ``off-state''. Similar behaviors can also be seen in the simulated output spectra for the ``on-state'' (Fig. 3d upper panel) and the ``off-state'' (Fig. 3d lower panel). In the Supplementary Information Section VI, we provide a detailed analysis of the nonlinear dynamics that explains the origin of the spectral dips. We also explain how the spectral dips are indicative of the power flow direction. It is worth noting that when the input pulse energy exceeds 600 fJ, the input FH exhibits significant spectral broadening during the SHG\cite{Marc_2018} and the phase relation between the FH and SH changes. This leads to a lower parametric gain of the DOPA (see Supplementary Information Section VI for details).

We further characterize the all-optical switching dynamics and the switching energy of the nonlinear splitter device using a degenerate pump–probe technique, similar to the measurement technique used in Ref.\cite{ono2020ultrafast}  As shown in Fig. 4a, the beam containing \textasciitilde46-fs-long pulses centered at 2090 nm generated from a table-top degenerate optical parametric oscillator\cite{marandi_cascaded_2016} is split into two beams by a beam splitter in a Michelson interferometer. One beam with a weak optical fluence (3 fJ, 770 nW on chip average power) is used as the probe beam, and another beam with a high/tunable optical fluence and adjustable time delay (controlled by a motorized delay stage) is used as the pump beam. In the measurement, we couple the output of the Michelson interferometer which contains both the pump and the probe beams to our chip, and switch the transmission of the probe signal by controlling the power of the pump pulse. To suppress the interferences between the pump and the probe beams, we use a piezoelectric transducer (PZT) in the optical path of the probe and modulate the phase of the probe pulses at 350 Hz, which is much faster than the measurement bandwidth. Additionally, we employ the lock-in modulation and demodulation scheme at 80 Hz to acquire the output probe signal only, rather than acquiring both the pump and probe signals. 

The dynamics of the ``on-state'' for the probe signal is measured at the through port and the drop port as plotted in Fig. 4 b and 4c, respectively. At the through port, we observe that the probe pulse is clearly appearing when the pump pulse temporally overlaps with it, whereas at the drop port the probe pulse is eliminated. To extract the rise and fall times of the switching mechanism, we fitted the data with exponential growth and decay functions for relative time delays  $\Delta\tau$ < 0 fs and $\Delta\tau$ > 0 fs, respectively, convolved with the autocorrelation of the input pulse, which was approximated by a Gaussian profile with a FWHM of 65.2 fs. (see Supplementary Information II for details)  For the measurement at the through port (Fig. 4b), the best fit yields a switching rise time of (5.1 ± 1) fs and a fall time of (18.1 ± 2.9) fs. The measurement at the drop port shows a switching rise time of (1.75 ± 0.86) fs and a fall time of (11.0 ± 2.15) fs. The measured ultrafast dynamics confirms that the all-optical switching mechanism is indeed the result of instantaneous quadratic nonlinearity without the contribution of other slow nonlinearities. However, the entire all-optical switch still has a finite response time due to the non-zero GVM, GVD and the finite phase matching bandwidth, which make switching for pulses shorter than 20 fs challenging. We have confirmed with numerical simulations that the asymmetry in the switching dynamics could be caused by the interplay of several dispersion mechanisms such as the GVM and the GVD. 

The measured ultrafast switching also indicates that the slow carrier dynamics that are commonly observed in LN and other materials, such as photorefractive effect\cite{jiang2017fast} and photothermal effect\cite{ryckman2012photothermal} are absent, primarily due to the low photon energies of the input FH and the generated SH, the ultra-low input pulse energy and the non-resonant nature of the switch. We have also ruled out the intensity-dependent index change (Kerr effect) of the LN waveguide as the possible mechanism for the switching and nonlinear transmission characteristics\cite{finlayson1988picosecond}, given the fact that no switching behavior and power-dependent transmission was observed on a similar device without poling in the main waveguide. 

Figure 4d and e show the extinction ratio between the ``on-'' and ``off-'' states of the switch for both output ports at $\Delta\tau$ = 0. We estimated a switching pump energy of 80 fJ (20 fJ) for the through port (the drop port) at the 3-dB contrast level. Within 500 fJ of input pulse energy, we obtain a switching contrast over 5 dB in the through port and a switching contrast over 8 dB in the drop port. Such a high switching contrast is difficult to realize using the saturable absorption in semiconductors\cite{spuhler2005semiconductor} and low-dimensional materials\cite{li2014ultrafast,bao2009atomic,set2004laser} without coupling them to optical cavities, since it, in general, requires very high optical fluence to excite a significant portion of electrons from the valence band to the conduction band.

In summary, we demonstrate on-chip all-optical switching based on the strong instantaneous nonlinear response of an LN nano-waveguide, with simultaneous dispersion and QPM engineering. Our on-chip device features the state-of-the-art switching energy-time product of $1.4 \times10^{-27}$ J$\cdot$s, which is an order of magnitude improvement over the previous all-optical switch based on graphene-loaded plasmonic waveguides\cite{ono2020ultrafast}. This nonlinear splitter also enables the simultaneous realizations of switch-on and -off operations as well as a large extinction ratio of over 5 dB when subjected to less than 500 fJ of input pulse energy. Moreover, due to the exceptionally low GVM and GVD of our dispersion-engineered LN waveguide, we can further improve the performance limit of the device by simply prolonging the poled region in the main waveguide. Our numerical simulation (see Supplementary Information section VII for details) has revealed that with the same dispersion characteristics and  input pulse configuration of 45 fs at 2.09 $\mu$m, a 70-mm-long-main waveguide (containing a 20-mm-long SHG and a 50-mm-long DOPA regions) yields a 2 dB switching contrast at a switching energy of 800 aJ. We envision that the low required switching energy (on the order of 10s of fJ) and the THz bandwidth of our all-optical switch make it amenable to directly interfacing with recently developed LN-based on-chip pulsed light sources, such as Kerr soliton micro-combs\cite{he2019self,gong2020near,gong2019soliton} and electro-optical (EO) combs\cite{zhang2019broadband}, towards on-chip ultrafast information processing\cite{bogaerts2020programmable} and time-multiplexed photonic networks\cite{leefmans2021topological}.

\section*{Methods}
\noindent \textbf{Device fabrication.} We fabricated the nonlinear splitter devices on a 700-nm-thick X-cut MgO-doped LN thin-film on 2-$\mu$m-thick SiO$_2$ on top of a LN substrate (NANOLN). We first patterned the poling electrodes (15 nm Cr/55 nm Au) with varied electrode finger periodicities using e-beam lithography. Then the electrodes were formed by e-beam evaporation and metal lift-off. We performed the ferroelectric domain inversion (periodic poling) by applying several 380 V, 5-ms-long pulses at room temperature with the sample submerged in oil. We visually inspected the poling quality using a second harmonic microscope. Next, we removed the electrodes by wet chemical etching, and patterned the waveguides using the e-beam lithography. The pattern was transferred to the LN layer by Ar$^+$ plasma etching. Finally, the waveguide facets were polished to enable good light coupling efficiencies.\\

\noindent \textbf{Optical measurements.} For the linear and nonlinear optical measurements, we employed a free-space light coupling setup shown in Fig. 4a. The 1045 nm source is a 1 W Yb mode-locked laser that produces nearly transform-limited 75-fs-long pulses at a 250 MHz repetition rate (Menlo Systems Orange). The output 1045 nm beam was fed into a near-synchronously pumped degenerate OPO\cite{Marc_2018} to produce $\sim46$-fs-long pulses centered at 2090 nm. The detailed characterization of the 2090 nm pulses is discussed in the Supplementary Information Section II.  The output 2090 nm beam was split into two beams by a beam splitter in a Michelson interferometer. Then the two beams were recombined and coupled into the nonlinear splitter chip by a reflective objective (Newport 50102-02). The average off-chip input power was calibrated by a thermal power meter (Thorlabs PM16-401). The input/output coupling losses at 2090 nm were estimated to be 21.6 dB/4 dB. For the power dependent transmittance measurements in Fig. 3, only one output beam from the Michelson interferometer was used. The chip was placed on a thermoelectric cooling stage (TEC). For adjusting the QPM condition, temperature tuning and thin organic materials were used (see Supplementary Information Section IV for details). For the results in Fig. 3 a and b, the output power was measured by an optical spectrum analyzer (OSA) covering 1200-2400 nm (Yokogawa AQ6375B) with a 2 nm resolution bandwidth. For the result in Fig. 4b-e, the output power was measured by an IR 2-$\mu$m photoreceiver (Newport 2034).\\

\noindent \textbf{Numerical simulations.} We used a commercial software (Lumerical Inc.) to solve for the waveguide modes as well as to obtain the dispersion characteristics shown in Fig. 2c. In the simulation, the anisotropic index of the LN was modeled by the Sellmeier equations\cite{zelmon1997infrared}. For the nonlinear optical simulation, we solved an analytical nonlinear envelope equation (NEE) in the frequency domain using a split-step Fourier technique to simulate the pulse propagation and nonlinear dynamics in the waveguide\cite{phillips_supercontinuum_2011}. The nonlinear step was solved with a fourth-order Runge-Kutta method. The details regarding the single-envelope simulation can be found in Supplementary Information Section III.


\section*{Data Availability}
The data that support the plots within this paper and other findings of this study
are available from the corresponding author upon reasonable request.
\section*{Code Availability}
The computer code used to perform the nonlinear simulations in this paper is available from the corresponding author upon reasonable request.
\section*{Acknowledgements}
The device nanofabrication was performed at the Kavli Nanoscience Institute (KNI) at Caltech. The authors thank Prof. Kerry Vahala and Prof. Changhuei Yang for loaning equipment. The authors gratefully acknowledge support from ARO grant no. W911NF-18-1-0285, NSF grant no. 1846273 and 1918549, AFOSR award FA9550-20-1-0040, and NASA/JPL. The authors wish to thank NTT Research for their financial and technical support.
\section*{Authors Contributions}
Q.G. and A.M. conceived the project; Q.G. fabricated the devices and performed the measurements with assistance from R.S, R.N., S.J. and R.M.G. L.L. developed the single-envelope simulation tool. L.L., D.J.D. and A.R. contributed to the design of the device; Q.G. and L.L. analyzed the experimental results and performed the simulations. L.L. performed the periodic poling; Q.G. wrote the manuscript with input from all other authors. A.M. supervised the project.
\section*{Competing Interests}
The authors declare no competing interests.

\newpage
\bibliography{references}

\end{document}